\documentclass[sigconf]{acmart}

\copyrightyear{2026}
\acmYear{2026}
\setcopyright{cc}
\setcctype{by}
\acmConference
    [SIGIR '26]
    {Proceedings of the 49th International ACM SIGIR Conference on Research and Development in Information Retrieval}
    {July 20--24, 2026}
    {Melbourne, VIC, Australia}
\acmBooktitle
    {Proceedings of the 49th International ACM SIGIR Conference on Research and Development in Information Retrieval (SIGIR '26), July 20--24, 2026, Melbourne, VIC, Australia}
\acmDOI{10.1145/3805712.3809560}
\acmISBN{979-8-4007-2599-9/2026/07}

\usepackage{enumitem}
\usepackage{amsmath}
\usepackage{multirow}
\usepackage{subcaption}

\begin{document}

\title
    [FLAME: Condensing Ensemble Diversity into a Single Network for Efficient Sequential Recommendation]
    {FLAME: Condensing Ensemble Diversity into a Single Network for Efficient Sequential Recommendation}

\author{WooJoo Kim}
\email{kimuj0103@postech.ac.kr}
\affiliation
{
    \institution{Pohang University of\\Science and Technology}
    \city{Pohang}
    \country{Republic of Korea}
}
\author{JunYoung Kim}
\email{junyoungkim@postech.ac.kr}
\affiliation
{
    \institution{Pohang University of\\Science and Technology}
    \city{Pohang}
    \country{Republic of Korea}
}
\author{JaeHyung Lim}
\email{jaehyunglim@postech.ac.kr}
\affiliation
{
    \institution{Pohang University of\\Science and Technology}
    \city{Pohang}
    \country{Republic of Korea}
}
\author{SeongJin Choi}
\email{sjin9805@postech.ac.kr}
\affiliation
{
    \institution{Pohang University of\\Science and Technology}
    \city{Pohang}
    \country{Republic of Korea}
}
\author{SeongKu Kang}
\authornote{Corresponding authors.}
\email{seongkukang@korea.ac.kr}
\affiliation
{
    \institution{Korea University}
    \city{Seoul}
    \country{Republic of Korea}
}
\author{HwanJo Yu}
\authornotemark[1]
\email{hwanjoyu@postech.ac.kr}
\affiliation
{
    \institution{Pohang University of\\Science and Technology}
    \city{Pohang}
    \country{Republic of Korea}
}
\renewcommand{\shortauthors}{WooJoo Kim et al.}

\begin{abstract}
    Sequential recommendation requires capturing diverse user behaviors, which a single network often fails to capture.
    While ensemble methods mitigate this, training multiple networks from scratch incurs high computational cost and instability from noisy mutual supervision.
    We propose \textbf{F}rozen and \textbf{L}earnable networks with \textbf{A}ligned \textbf{M}odular \textbf{E}nsemble (\textbf{FLAME}), a novel framework that condenses ensemble-level diversity into a single network for efficient sequential recommendation.
    During training, FLAME simulates exponential diversity using only two networks via \textit{modular ensemble}, which dynamically combines sub-modules (e.g., layers) of each network to generate a rich space of diverse representation patterns.
    To stabilize training, FLAME pretrains and freezes one network as a semantic anchor and employs \textit{guided mutual learning} to align diverse representations into the space of remaining learnable network.
    At inference, FLAME utilizes only the learnable network, achieving ensemble-level performance with zero overhead compared to a single network.
    Experiments on six datasets show that FLAME outperforms state-of-the-art baselines, achieving up to 7.69$\times$ faster convergence and 9.70\% improvement in NDCG@20.\footnote{Our code is available at \url{https://github.com/woo-joo/FLAME_SIGIR26}.}
\end{abstract}

\begin{CCSXML}
<ccs2012>
   <concept>
       <concept_id>10002951.10003317.10003347.10003350</concept_id>
       <concept_desc>Information systems~Recommender systems</concept_desc>
       <concept_significance>500</concept_significance>
       </concept>
   <concept>
       <concept_id>10002951.10003317.10003331.10003271</concept_id>
       <concept_desc>Information systems~Personalization</concept_desc>
       <concept_significance>500</concept_significance>
       </concept>
   <concept>
       <concept_id>10002951.10003260.10003261.10003269</concept_id>
       <concept_desc>Information systems~Collaborative filtering</concept_desc>
       <concept_significance>500</concept_significance>
       </concept>
 </ccs2012>
\end{CCSXML}

\ccsdesc[500]{Information systems~Recommender systems}
\ccsdesc[500]{Information systems~Personalization}
\ccsdesc[500]{Information systems~Collaborative filtering}

\keywords{Sequential Recommendation, Modular Ensemble, Guided Mutual Learning}

\maketitle

\section{Introduction}

Sequential recommendation presents non-trivial challenges due to diverse user behaviors, complex item spaces, and multi-faceted interactions \cite{hidasi2015session, kang2018self, rendle2010factorizing, tang2018personalized, chen2020sequence, wu2019session, xie2022contrastive, wang2015learning, lee2026capturing, hwang2024multi}.
A single user sequence can further reflect various patterns--such as short-term interests, long-term preferences, and shifts in user tastes--making it difficult for a single model to capture all relevant signals \cite{wang2015learning, ma2020disentangled, qiu2020exploiting, qiu2019rethinking, lim2025federated, cho2023dynamic}.
To interpret sequences from multiple perspectives, a promising direction is combining multiple models, each focusing on complementary aspects of the data.
This core idea drives ensemble methods \cite{fort2019deep, allen2020towards, du2023ensemble}, which have shown strong performance across domains by capturing diverse and complementary patterns.
In this light, ensemble learning is particularly well-suited for sequential recommendation, where user intents and item semantics vary widely.

Recently, several ensemble-based methods have emerged to capture the diverse dynamics in user sequences \cite{du2023sequential, kang2023distillation, zhang2025frequency}.
These approaches commonly train multiple networks from scratch in parallel, encouraging each network to specialize in complementary patterns.
Inspired by recent advancements in computer vision \cite{song2018collaborative, zhu2018knowledge, zhang2018deep}, many of these methods \cite{du2023ensemble, zhang2025frequency} incorporate mutual knowledge transfer among networks to enhance collective learning.
For instance, EMKD \cite{du2023ensemble} leverages contrastive learning and knowledge distillation to align network outputs while preserving internal representation diversity.
Overall, these strategies promote inter-network agreement and improve optimization over training networks independently.

Despite their effectiveness, recent ensembles face two major limitations: \textit{computational inefficiency} and \textit{training instability}.
First, employing multiple networks significantly increases computational cost, as ensemble prediction requires forwarding each network independently and aggregating their outputs.
This makes training and inference several times more expensive than a single network, limiting practicality for real-time services.
Second, training all networks from scratch with mutual knowledge transfer often leads to unstable optimization, especially in the early stages when none of the networks has reliable prior knowledge \cite{li2021not, xie2022performance}.
This issue can be further aggravated in sequential recommendation, where data is often sparse and noisy \cite{han2018co}.
In our experiments, existing ensemble methods exhibit large training fluctuations and require up to 2.94$\times$ longer to converge than a single network (see Section~\ref{sec/efficiency_and_stability_study}).

To address the inefficiency and instability of ensemble methods, we focus on two key insights from prior literature and our empirical study.
First, \textit{different sub-modules within a single network can capture distinct patterns}.
Prior works \cite{dosovitskiy2016inverting, kulkarni2017layer, wang2023understanding} show deeper layers tend to produce increasingly abstract and discriminative representations.
A similar phenomenon can be expected in sequential recommendation: embedding modules capture static item semantics, early encoder layers reflect localized and item-specific signals, and deeper layers encode temporally smoothed user preferences.
This structural separation suggests a single network can internally generate diverse representations by interpreting sequences from multiple perspectives.
Building on this, we simulate ensemble-like behavior via exponentially many sub-module combinations from only two networks, without explicitly building multiple networks.

\begin{figure}[t]
  \centering
  \begin{subfigure}[t]{0.23\textwidth}
    \centering
    \includegraphics[width=\textwidth]{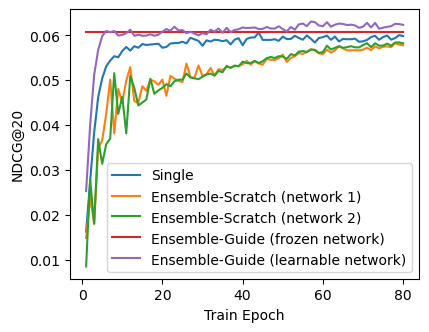}
    \caption{Beauty}
  \end{subfigure}
  \hfill
  \begin{subfigure}[t]{0.23\textwidth}
    \centering
    \includegraphics[width=\textwidth]{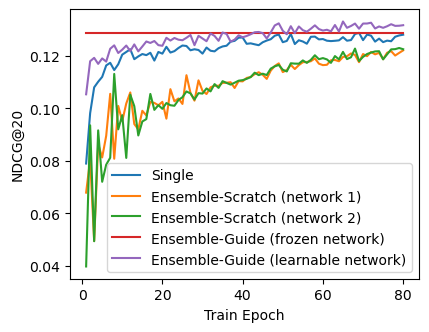}
    \caption{MovieLens 1M}
  \end{subfigure}
  \caption{Training curves for Single, Ensemble-Scratch, and Ensemble-Guide.}
  \label{fig/ensemble_instability}
  \Description{}
\end{figure}

Second, \textit{pretraining and freezing one network of an ensemble can substantially improve training stability}.
We empirically observe that a single pretrained network, trained identically but without mutual knowledge transfer, provides effective prior knowledge for the remaining learnable networks, ensuring stable and effective optimization.
In \autoref{fig/ensemble_instability}, we evaluate SASRec \cite{kang2018self} under three settings: Single, Ensemble-Scratch (two SASRecs trained from scratch with mutual knowledge transfer\footnote{Mutual knowledge transfer in Ensemble-Scratch--implemented via contrastive learning and knowledge distillation--is analogous to the approach used in EMKD \cite{du2023ensemble}.}), and Ensemble-Guide (one network in Ensemble-Scratch is pretrained and frozen).
Ensemble-Scratch networks exhibit larger fluctuations and slower convergence than Single, underperforming due to noisy supervision between untrained networks.
Conversely, the learnable network in Ensemble-Guide converges faster and more stably, outperforming Single via guidance from the frozen network.
This aligns with self-distillation studies, where a pretrained teacher improves training stability and performance of an identical student \cite{furlanello2018born, pham2022revisiting, pareek2024understanding}.

Building on these insights, we propose \textbf{F}rozen and \textbf{L}earnable networks with \textbf{A}ligned \textbf{M}odular \textbf{E}nsemble (\textbf{FLAME}), a novel framework that condenses ensemble diversity into a single network for efficient sequential recommendation.
FLAME consists of two identical networks: one pretrained and frozen, and the other learned from scratch.
Each network comprises $M$ sub-modules, and each input traverses a combination of frozen and learnable sub-modules, yielding $2^M$ unique representations.
These representations are aligned into a unified semantic space via contrastive learning, utilizing the frozen network as a stable semantic anchor.
This design allows FLAME to simulate the diversity of an exponentially large ensemble using only two networks, enabling the learnable network to stably capture complementary patterns.
At inference, FLAME uses only the learnable network, achieving ensemble-level performance with the single-network efficiency.
Thus, FLAME provides a practical and robust alternative to conventional ensemble methods.

Our main contributions are as follows:
\begin{itemize}[itemsep=1pt, left=0pt]
    \item We provide both conceptual and empirical evidence revealing the inefficiency and instability of conventional ensemble.
    \item We propose FLAME, a framework that condenses ensemble diversity into a single network, achieving robust training stability and zero-overhead inference efficiency.
    \item We conduct extensive experiments on six real-world datasets, where FLAME consistently outperforms state-of-the-art baselines with significantly lower training cost and inference latency.
\end{itemize}

\section{Preliminary and Related Work}

\subsection{Problem Formulation}

In the sequential recommendation setting, the goal is to predict which item a user will interact with next, given her chronological sequence of past interactions. Let $\mathcal{U}$ and $\mathcal{I}$ denote the sets of users and items, respectively. Each user $u \in \mathcal{U}$ is associated with an ordered sequence of items $s_u = [i_{u, 1}, i_{u, 2}, \ldots, i_{u, |s_u|}]$, where $i_{u, t} \in \mathcal{I}$ indicates the $t$-th item that user $u$ has interacted with. Based on the observed sequence $s_u$, the sequential recommendation task aims to predict the next item $i_{u, |s_u|+1}$ that user $u$ is likely to interact with, which can be formalized as:
\begin{equation}
    \underset{i \in \mathcal{I}}{\operatorname{argmax}} \ p(i_{u, |s_u|+1} = i \mid s_u).
\end{equation}

\begin{figure}[t]
    \centering
    \includegraphics[width=\linewidth]{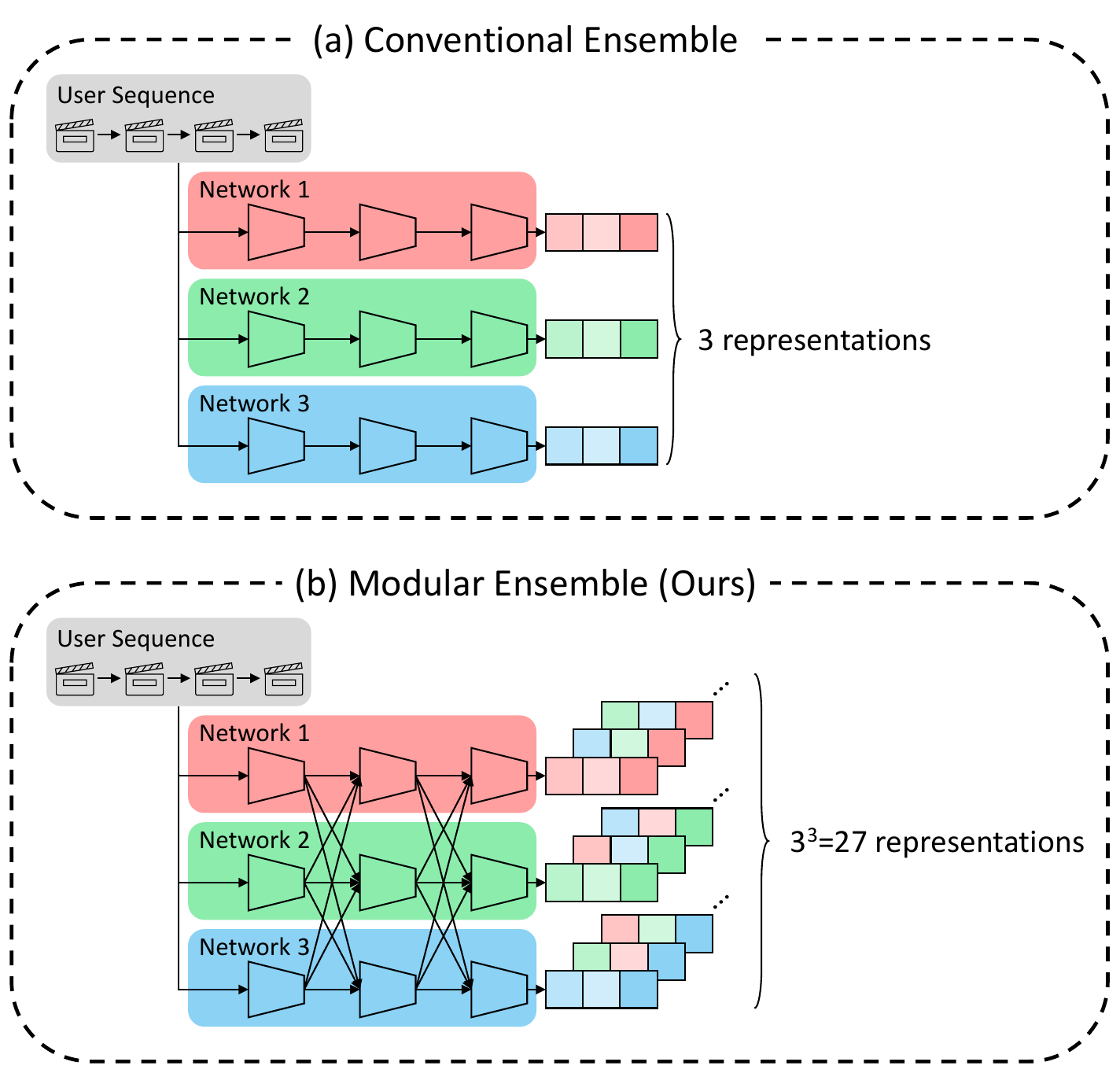}
    \caption{Conceptual illustration of (a) conventional ensemble and (b) proposed modular ensemble. With $N$ networks, conventional ensemble produces $N$ distinct representations. When each network is decomposed into $M$ sub-modules, modular ensemble generates $N^M$ different representations.}
    \label{fig/modular_ensemble}
    \Description{}
\end{figure}

\subsection{Ensemble-based Sequential Recommender}

Ensemble learning is a well-established machine learning technique that combines the outputs of multiple networks to enhance prediction accuracy and generalization performance \cite{breiman1996bagging, freund1997decision, dietterich2000ensemble}.
In the context of sequential recommendation, ensemble methods can be applied to capture diverse user behavior patterns, thereby enhancing top-k recommendation accuracy.
Concretely, a typical ensemble-based sequential recommender constructs $N$ separate networks and forwards each user sequence $s_u$ through all of them in parallel, as illustrated in \autoref{fig/modular_ensemble}a. This results in a set of representations $\{ \tilde{h}_u^n \}_{n=1}^N$, where $n$ indexes each network. The model is then trained using a composite objective function:
\begin{equation}
    \mathcal{L} = \sum_{u \in \mathcal{U}} \left( \sum_{1 \le n \le N} \mathcal{L}_{rec}\left(u;\tilde{h}_u^n\right) + \mathcal{L}_{mkt}\left(u;\tilde{h}_u^1, \tilde{h}_u^2, \ldots, \tilde{h}_u^N\right) \right).
\end{equation}
Here, $\mathcal{L}_{rec}$ ensures that each network individually learns the sequential recommendation task, while $\mathcal{L}_{mkt}$ facilitates mutual knowledge transfer among networks through pairwise objectives:
\begin{equation}
    \mathcal{L}_{mkt} = \sum_{1 \le n_1, n_2 \le N, n_1 \ne n_2} \mathcal{L}_{pair}(\tilde{h}_u^{n_1}, \tilde{h}_u^{n_2}).
\end{equation}

Recent ensemble-based approaches adopt this training paradigm.
For instance, EMKD \cite{du2023ensemble} ensembles BERT4Recs \cite{sun2019bert4rec} with knowledge distillation, SEM \cite{du2023sequential} adopts multiple networks for diversity-aware learning, and FamouSRec \cite{zhang2025frequency} leverages heterogeneous architectures to capture different frequency patterns via contrastive alignment.
While network ensembling has also been studied in click-through rate prediction \cite{zhu2020ensembled, zhang2022dhen, liu2024collaborative}, its goal there is to handle diverse external features rather than varied patterns within ID-based user sequences.
Despite their effectiveness, conventional ensemble-based sequential recommenders require explicitly constructing multiple networks to ensure diversity, incurring substantial computational cost.
Moreover, mutual knowledge transfer among randomly initialized networks, without reliable prior knowledge, can destabilize training and lead to sub-optimal performance.

\subsection{Backbone Network}

Motivated by the capability of Transformer to model intricate sequential patterns \cite{vaswani2017attention, devlin2019bert}, FLAME employs SASRec \cite{kang2018self} as the backbone for each network.
A wide range of recent studies \cite{zhou2020s3, qin2023meta, liu2025facet, zhang2025frequency} also adopt SASRec as their base architecture, including contrastive learning \cite{liu2021contrastive, dang2023uniform, wei2023moco4srec, wang2024relative} and intent-aware \cite{chen2022intent, qin2024intent, li2023multi, wang2025intent} approaches, demonstrating its versatility and strong representational capacity across learning paradigms.
SASRec comprises two primary components: an embedding module and a sequence encoding module.

\subsubsection{Embedding Module}
The network maintains two learnable embedding tables: an item embedding matrix $I \in \mathbb{R}^{|\mathcal{I}| \times d}$ and a position embedding matrix $P \in \mathbb{R}^{T \times d}$, where $T$ is the maximum sequence length. For a given sequence $s_u$, its length is first adjusted to $T$ by either truncating to the most recent $T$ items or applying zero padding at the beginning. The sequence embedding $e_u$ is then constructed as follows:
\begin{equation}
    e_u = [I_{i_{u, 1}} + P_1, I_{i_{u, 2}} + P_2, \ldots, I_{i_{u, T}} + P_T] \in \mathbb{R}^{T \times d},
\end{equation}
where $I_{i_{u, t}} \in \mathbb{R}^d$ and $P_t \in \mathbb{R}^d$ correspond to the embeddings of item $i_{u, t}$ and position $t$, respectively.

\subsubsection{Sequence Encoding Module}
The resulting embedding $e_u$ is then processed by a $L$-layer multi-head Transformer encoder whose $l$-th layer is denoted as $\text{Trm}^l(\cdot)$. With $H_u^0 = e_u$, each layer yields:
\begin{equation}
    H_u^l = [h_{u, 1}^l, h_{u, 2}^l, \ldots, h_{u, T}^l] = \text{Trm}^l(H_u^{l-1}) \in \mathbb{R}^{T \times d}.
\end{equation}
Here, the hidden representation corresponding to the final position in the output of the last layer, denoted as $h_{u, T}^L \in \mathbb{R}^d$, is utilized as the final representation of the user sequence $s_u$.
For brevity, we denote the final representation of user $u$ as $h_u$ hereafter.

\section{Design Principle}
\begin{figure*}[t]
  \centering
  \begin{subfigure}[t]{0.15\textwidth}
    \centering
    \includegraphics[width=\textwidth]{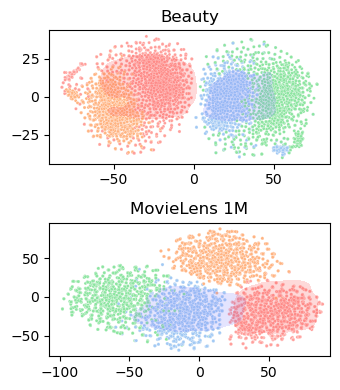}
    \caption{Distribution}
  \end{subfigure}
  \hfill
  \begin{subfigure}[t]{0.21\textwidth}
    \centering
    \includegraphics[width=\textwidth]{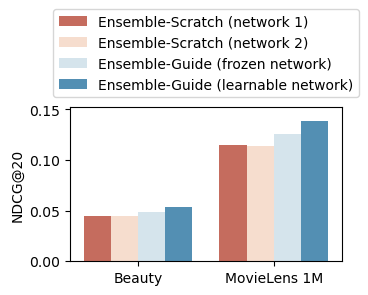}
    \caption{Performance}
  \end{subfigure}
  \hfill
  \begin{subfigure}[t]{0.28\textwidth}
    \centering
    \includegraphics[width=\textwidth]{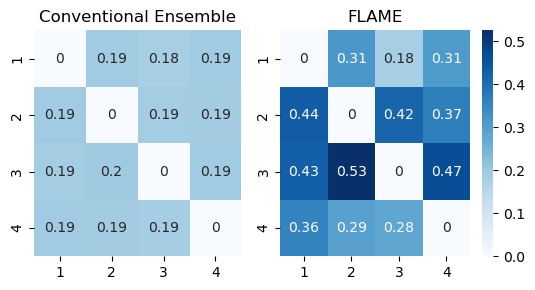}
    \caption{Beauty}
  \end{subfigure}
  \hfill
  \begin{subfigure}[t]{0.28\textwidth}
    \centering
    \includegraphics[width=\textwidth]{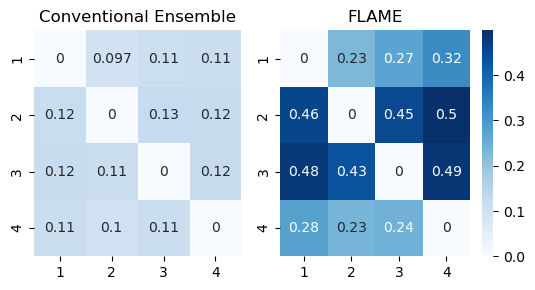}
    \caption{MovieLens 1M}
  \end{subfigure}
  \caption{(a) t-SNE visualization of sequence representations for conventional ensemble (red and blue shaded areas) and modular ensemble (red, yellow, green, and blue colored points). (b) Performance for individual network in Ensemble-Scratch and Ensemble-Guide. (c, d) PER maps for conventional ensemble and FLAME.}
  \label{fig/ensemble_diversity_reliability}
  \Description{}
\end{figure*}

Before detailing the proposed FLAME framework, we first elaborate on its two foundational designs--\textit{modular ensemble} and \textit{guided mutual learning}--which are designed to mitigate the computational inefficiency and training instability inherent in conventional ensemble approaches.

\subsection{Modular Ensemble}
Modular ensemble, depicted in \autoref{fig/modular_ensemble}b, addresses the computational inefficiency of conventional ensemble methods while preserving, or even enhancing, representational diversity.
Unlike conventional ensemble, which generates representations by forwarding the input through an entire network as a whole, modular ensemble flexibly chooses the sub-module to use at each stage from any of the networks.
This design allows for a combinatorial expansion of representation paths without increasing the number of complete networks.
For example, with three networks each split into three sub-modules, the modular ensemble produces $3^3 = 27$ unique representations, compared to only 3 in a conventional setup.
Conversely, even a small number of networks (e.g., two networks with three sub-modules each) can simulate the effect of an 8-network conventional ensemble through $2^3 = 8$ modular combinations, significantly improving efficiency.

\subsection{Guided Mutual Learning}
Another critical challenge of conventional ensemble training, as observed in \autoref{fig/ensemble_instability}, is optimization instability, presumably stemming from noisy supervision exchanged among randomly initialized networks.
This issue can be alleviated by pretraining and freezing some of the networks in ensemble.
The frozen networks, imbued with prior knowledge, serves as a stable reference during training, guiding the rest learnable networks toward faster and more reliable convergence.
In addition to stabilizing, this setup also improves the standalone quality of the learnable networks through mutual enhancement between networks \cite{zhang2018deep, kweon2021bidirectional, xie2022performance}.
This strategy integrates particularly well with modular ensemble, where fine-grained knowledge exchange occurs structurally at the sub-module level.

\subsection{Empirical Evidence}
To motivate the integration of the two core designs into FLAME framework, we empirically examine their effectiveness in the context of ensemble learning.
For an ensemble to be effective, it must satisfy two critical conditions: \textit{diversity} among representations and \textit{reliability} of each individual network \cite{dietterich2000ensemble, zhou2025ensemble}.
From this perspective, we evaluate how the introduced designs--modular ensemble and guided mutual learning--are specifically tailored to achieve diversity and reliability, respectively, and thus serve as a solid foundation for an ensemble recommender.

\textbf{Diversity.}
\autoref{fig/ensemble_diversity_reliability}a visualizes the distribution of sequence representations generated by two distinct SASRec-based ensemble strategies: 1) a 2-network conventional ensemble and 2) a 2-network modular ensemble, where each network is decomposed into two sub-modules.
While both methods produce separated representations, modular ensemble doubles the output diversity (four cluster types versus two) for the same number of networks.
Specifically, because different sub-modules within a single network capture distinct patterns, their combinations enable the creation of an exponentially diverse set of representations.
Consequently, modular ensemble expands its representational coverage to include the diverse user behaviors (i.e., the yellow and green clusters) entirely missed by conventional ensemble.
This capacity for capturing the multifaceted nature of user interests is critical for advancing sequential recommendation performance.

\textbf{Reliability.}
\autoref{fig/ensemble_diversity_reliability}b presents the recommendation performance of individual network in Ensemble-Scratch and Ensemble-Guide in \autoref{fig/ensemble_instability}.
The learnable network in Ensemble-Guide achieves the highest performance among all four networks, indicating the guidance from the frozen network during training enhances the reliability of individual network.
It is also noteworthy that the two networks in Ensemble-Scratch both underperform compared to the frozen network in Ensemble-Guide (or equivalently, a single SASRec), suggesting that noisy mutual supervision among networks in the early stages of training can lead to sub-optimal convergence.

\textbf{Put it together.}
To validate the synergy between modular ensemble and guided mutual learning, we compare our proposed FLAME against a 4-network conventional ensemble, both built upon SASRec backbone.
Both methods produce four representations, and we evaluate their diversity and reliability via the pairwise exclusive-hit ratio (PER) \cite{kang2022consensus}, as shown in \autoref{fig/ensemble_diversity_reliability}c,d.
A higher PER signifies the capture of more beneficial, complementary patterns.\footnote{$\text{PER}(h_i, h_j) = \frac{|\mathcal{H}_i - \mathcal{H}_j|}{|\mathcal{H}_i|}$ where $\mathcal{H}_i$ denotes the set of users hit by $h_i$.}
Here, a hit is recorded for a user when her ground-truth target item is ranked within the top-20.
FLAME consistently records higher PER values, which demonstrates the successful fusion of its components: modular ensemble excels at capturing diverse user behavioral patterns, while guided mutual learning ensures these patterns are meaningful and vital for successful sequential recommendations.

\section{Proposed Framework: FLAME}
\begin{figure}[t]
    \centering
    \includegraphics[width=\linewidth]{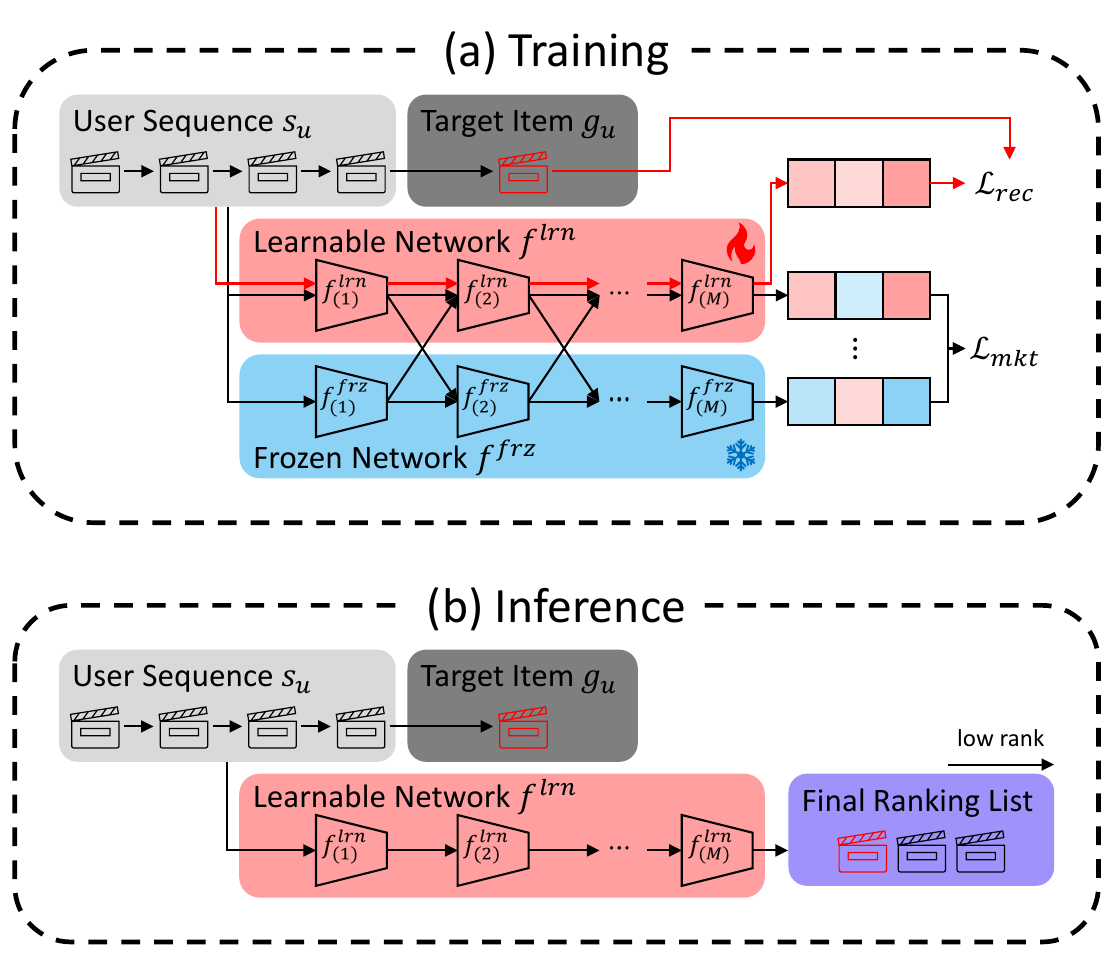}
    \caption{Illustration of (a) training and (b) inference procedure of proposed FLAME. In (a), the {\it Learnable} network is optimized via the next-item prediction task. In parallel, $2^M$ diverse representations are generated by modular ensemble of the {\it Frozen} and {\it Learnable} networks and then aligned into a unified semantic space. In (b), only the {\it Learnable} network is utilized to enable efficient inference.}
    \label{fig/flame}
    \Description{}
\end{figure}

We introduce \textbf{F}rozen and \textbf{L}earnable networks with \textbf{A}ligned \textbf{M}odular \textbf{E}nsemble for sequential recommendation, named \textbf{FLAME}, a unified framework combining modular ensemble with guided mutual learning, as illustrated in \autoref{fig/flame}.
FLAME utilizes only two networks, yet efficiently achieves representation diversity via modular ensemble.
One network is pretrained and frozen to enhance the reliability of the remaining learnable network through guided mutual learning.
FLAME performs accurate and efficient recommendation using only the learnable network at inference time.

\subsection{Ensemble Construction} \label{sec/ensemble_construction}

\subsubsection{Network and Sub-Module Configuration}
FLAME constructs a modular ensemble of $N = 2$ networks: a pretrained \textit{Frozen} network, denoted as $f^{frz}(\cdot)$, and a \textit{Learnable} network, $f^{lrn}(\cdot)$, which is trained from scratch.
To encourage diversity, each network is further decomposed into $M$ sub-modules, with $f^{frz}_{(i)}(\cdot)$ and $f^{lrn}_{(i)}(\cdot)$ denoting the $i$-th sub-modules of the \textit{Frozen} and \textit{Learnable} networks.
For given user sequence $s_u$, it is processed sequentially through the sub-modules from the first to the $M$-th, with the final sub-module output serving as the resulting representation:
\begin{equation}
    \begin{aligned}
        f^{frz}(s_u) & = \left( f^{frz}_{(M)} \circ f^{frz}_{(M-1)} \circ \ldots \circ f^{frz}_{(1)} \right) (s_u) = h_u^{frz}, \\
        f^{lrn}(s_u) & = \left( f^{lrn}_{(M)} \circ f^{lrn}_{(M-1)} \circ \ldots \circ f^{lrn}_{(1)} \right) (s_u) = h_u^{lrn}.
    \end{aligned}
\end{equation}

\textbf{Instantiation.}
In our implementation, we partition each network into $M = 2$ sub-modules: an embedding module and a sequence encoding module.
While increasing $M$ could bring higher diversity, we empirically obtain satisfactory results with $M = 2$.
Further investigation of finer-grained sub-module decompositions (i.e., larger $M$) is discussed in Section~\ref{sec/scalability_analysis}.

\subsubsection{Sub-Module Combination}
FLAME constructs a rich representation space from only two networks by combining their sub-modules to capture diverse behavioral patterns.
At each of the $M$ sub-module stages, it selects a component from either the \textit{Frozen} or \textit{Learnable} network, yielding $2^M$ distinct computational paths.
Formally, each representation is defined as:
\begin{equation}
    \left( f^{p_M}_{(M)} \circ f^{p_{M-1}}_{(M-1)} \circ \ldots \circ f^{p_1}_{(1)} \right) (s_u) = \tilde{h}_u^p,
\end{equation}
where $p = (p_1, p_2, \ldots, p_M) \in \mathcal{P} = \{ frz, lrn \}^M$ specifies the decision path across sub-module stages.
Each $p_i \in \{ frz, lrn \}$ indicates whether the $i$-th sub-module is taken from the \textit{Frozen} ($frz$) or \textit{Learnable} ($lrn$) network.
Thus, each sequence representation $\tilde{h}_u^p$ corresponds to a unique path via modular ensemble.

\subsection{Mutual Knowledge Transfer}
FLAME aligns the numerous representations from its sub-module combinations into a unified semantic space for guided mutual learning.
This alignment serves two purposes: improving the reliability of learned representations \cite{liu2019deep, fort2019deep} and leveraging the \textit{Frozen} network to provide stable guidance during training \cite{allen2020towards}.

To achieve this, we adopt a pairwise contrastive learning objective where representations derived from the same user sequence are considered as positives, while those generated from different users in the same mini-batch act as negatives \cite{xie2022contrastive, qiu2022contrastive}.
Concretely, the alignment procedure is formulated as:
\begin{equation}
    \mathcal{L}_{cl}(u) = \sum_{\{ p, q \} \subset \mathcal{P}} \mathcal{L}_{nce}(\tilde{h}^p_u, \tilde{h}^q_u)
\end{equation}
and a noise contrastive estimation objective \cite{oord2018representation, liu2021contrastive} for each pair can be computed as follows:
\begin{equation}
    \mathcal{L}_{nce}(\tilde{h}^p_u, \tilde{h}^q_u) = -\log \frac{\exp(\tilde{h}^p_u \cdot \tilde{h}^q_u / \tau)}{\sum_{s_{u'} \in B} \exp(\tilde{h}^p_{u} \cdot \tilde{h}^q_{u'} / \tau)},
\end{equation}
where $\tau$ is the temperature parameter, and $B$ is the mini-batch of sequences.

\textbf{Similarity-based Weighting.}
The $2^M$ modular representations are not independent, as pairs sharing more sub-modules are inherently more similar due to their overlapping computational paths.
Consequently, a standard contrastive objective that treats all pairs as equally informative becomes inefficient, with the training loss being disproportionately dominated by these structurally similar pairs.

To mitigate this imbalance, we modify $\mathcal{L}_{cl}$ to dynamically adjust the weight assigned to each pair based on their similarity:
\begin{equation}
    \mathcal{L}_{mkt}(u) = \sum_{\{ p, q \} \subset \mathcal{P}} w_{p, q} \cdot \mathcal{L}_{nce}(\tilde{h}^p_u, \tilde{h}^q_u)
\end{equation}
and the similarity-based weight between each pair can then be formulated as follows:
\begin{equation}
    \begin{aligned}
        sim_{p, q} & = \frac{1}{B} \sum_{s_u \in B} \tilde{h}^p_u \cdot \tilde{h}^q_u, \\
        w_{p, q} & = \frac{\exp(-sim_{p, q})}{\sum_{\{ p', q' \} \subset \mathcal{P}} \exp(-sim_{p', q'})}.
    \end{aligned}
\end{equation}
Specifically, $\mathcal{L}_{mkt}$ down-weights pairs that are more similar, preventing them from dominating the overall objective and thereby promoting more balanced optimization.

\subsection{Optimization} \label{sec/optimization}
To endow the \textit{Learnable} network with the ability to directly perform sequential recommendation, we incorporate a next-item prediction objective into the training process of FLAME:
\begin{equation} \label{eq/L_rec}
    \begin{aligned}
        \hat{y}_u & = \text{Softmax}(h_u^{lrn} (I^{lrn})^\top) \in \mathbb{R}^{|\mathcal{I}|}, \\
        \mathcal{L}_{rec}(u) & = -\log \hat{y}_u[g_u],
    \end{aligned}
\end{equation}
where $g_u$ denotes the ground-truth item index for a user $u$.\footnote{The \textit{Frozen} network is pretrained using only $\mathcal{L}_{rec}$.}
In parallel, the contrastive alignment objective enhances the reliability and diversity of this representation.
To integrate these two complementary learning signals, we jointly optimize $\mathcal{L}_{rec}$ and $\mathcal{L}_{mkt}$ under a multi-task training strategy:
\begin{equation}
    \begin{aligned}
        \mathcal{L} & = \sum_{u \in \mathcal{U}} \left( \mathcal{L}_{rec}(u) + \lambda \cdot \mathcal{L}_{mkt}(u) \right),
    \end{aligned}
\end{equation}
where $\lambda$ controls the intensity of $\mathcal{L}_{mkt}$.

\textbf{Coefficient Annealing.}
To prevent the \textit{Learnable} network from overfitting to the representation space shaped by the \textit{Frozen} network, we gradually anneal the weighting coefficient $\lambda$ as training progresses:
\begin{equation}
    \lambda(r) = \lambda_0 \left( \frac{\lambda_R}{\lambda_0} \right)^\frac{r}{R},
\end{equation}
where $\lambda(r)$ denotes the coefficient of contrastive objective at epoch $r$, annealed from $\lambda_0$ and to $\lambda_R$ over $R$ total training epochs.
This scheduling allows the \textit{Learnable} network to benefit from the guidance of contrastive alignment in early phase, while gradually shifting focus toward improving its own task-specific representation quality.

\subsection{Single-Network Inference}
Nearly all modular representations\footnote{One exception constructed solely from the \textit{Frozen} sub-modules.} are constructed using at least one sub-module from the \textit{Learnable} network, and these are subsequently aligned into a unified semantic space.
As a result, the \textit{Learnable} network is implicitly trained to capture this entire space, thereby acquiring both the diversity and reliability induced by the ensemble learning.

This enables efficient inference: since the \textit{Learnable} network alone can approximate ensemble-level performance, we perform next-item prediction using only the \textit{Learnable} network:
\begin{equation}
    p(i_{u,n+1}|s_u) = \text{Softmax}(h_u^{lrn} I^\top) \in \mathbb{R}^{|\mathcal{I}|}.
\end{equation}
Inference remains as efficient as the original backbone (i.e., a single SASRec), incurring no additional cost.

\section{Experiment}
We seek to investigate the following research questions (RQs):

\begin{itemize}[itemsep=1pt, left=0pt]
    \item \textbf{RQ1}: How does FLAME perform compared to state-of-the-art baselines in sequential recommendation? (Section~\ref{sec/performance_comparison})
    \item \textbf{RQ2}: How effective are the core components of FLAME? (Section~\ref{sec/ablation_study})
    \item \textbf{RQ3}: How does FLAME scale with the number of sub-modules $M$? (Section~\ref{sec/scalability_analysis})
    \item \textbf{RQ4}: How robust is FLAME to different choices of the \textit{Frozen} network? (Section~\ref{sec/frozen_network_robustness})
    \item \textbf{RQ5}: Does FLAME solve computational inefficiency and training instability issues of previous ensemble methods? (Section~\ref{sec/efficiency_and_stability_study})
    \item \textbf{RQ6}: How do hyperparameters impact the performance of FLAME? (Section~\ref{sec/parameter_sensitivity})
\end{itemize}

\subsection{Experimental Setup}

\subsubsection{Dataset}
\begin{table}[t]
    \footnotesize
    \setlength{\tabcolsep}{3pt}
    \caption{Statistics of the datasets.}
    \label{tab/dataset}
    \begin{tabular}{c|ccccc}
        \toprule
        \textbf{Datasets} & \textbf{\# Interactions} & \textbf{\# Users} & \textbf{\# Items} & \textbf{Avg.SeqLen} & \textbf{Sparsity} \\
        \midrule
        Toys         & 167,597 & 19,412 & 11,924 &   8.6 & 99.93 \% \\
        Beauty       & 198,502 & 22,363 & 12,101 &   8.9 & 99.93 \% \\
        Games        & 231,780 & 24,303 & 10,672 &   9.5 & 99.91 \% \\
        Sports       & 296,337 & 35,598 & 18,357 &   8.3 & 99.96 \% \\
        Yelp         & 345,186 & 29,915 & 21,572 &  11.5 & 99.95 \% \\
        MovieLens 1M & 999,611 &  6,040 &  3,416 & 165.5 & 95.16 \% \\
        \bottomrule
    \end{tabular}
\end{table}

\begin{table*}[t]
    \scriptsize
    \setlength{\tabcolsep}{3.5pt}
    \caption{Model performance of FLAME and baselines on six public datasets. The bold score in each row represents the best and the underlined score represents the second-best. $Improv.$ denotes the relative improvement of FLAME against the best baseline.}
    \label{tab/performance}
    \begin{tabular}{c|c|ccccc|cccc|ccc||c|c}
        \toprule
        \textbf{Dataset} & \textbf{Metric} & \textbf{GRU4Rec} & \textbf{Caser} & \textbf{SASRec} & \textbf{BERT4Rec} & \textbf{FMLPRec} & \textbf{CL4SRec} & \textbf{DuoRec} & \textbf{MCLRec} & \textbf{CT4Rec} & \textbf{EMKD} & \textbf{DHEN} & \textbf{FamouSRec} & \textbf{FLAME} & $\boldsymbol{Improv.}$ \\
        \midrule\midrule
        \multirow{6}{*}{Toys}
                                      & HR@5    & 0.0316 & 0.0146 & 0.0554 & 0.0362 & 0.0620 & \underline{0.0707} & 0.0694 & 0.0692 & 0.0683 & 0.0598 & 0.0668 & 0.0682 & \bf{0.0744} &  5.21\% \\
                                      & HR@10   & 0.0496 & 0.0273 & 0.0737 & 0.0533 & 0.0855 & \underline{0.0952} & 0.0917 & 0.0946 & 0.0922 & 0.0925 & 0.0932 & 0.0912 & \bf{0.1026} &  7.82\% \\
                                      & HR@20   & 0.0746 & 0.0443 & 0.0989 & 0.0654 & 0.1140 & \underline{0.1268} & 0.1211 & 0.1257 & 0.1221 & 0.1256 & \underline{0.1268} & 0.1226 & \bf{0.1382} &  9.01\% \\
                                      & NDCG@5  & 0.0213 & 0.0102 & 0.0404 & 0.0246 & 0.0444 & 0.0501 & \underline{0.0502} & 0.0501 & 0.0497 & 0.0405 & 0.0487 & 0.0493 & \bf{0.0543} &  8.07\% \\
                                      & NDCG@10 & 0.0271 & 0.0142 & 0.0463 & 0.0300 & 0.0520 & 0.0580 & 0.0574 & \underline{0.0583} & 0.0573 & 0.0510 & 0.0552 & 0.0567 & \bf{0.0634} &  8.75\% \\
                                      & NDCG@20 & 0.0334 & 0.0185 & 0.0527 & 0.0364 & 0.0592 & 0.0660 & 0.0648 & \underline{0.0662} & 0.0648 & 0.0611 & 0.0643 & 0.0647 & \bf{0.0723} &  9.29\% \\
        \hline
        \multirow{6}{*}{Beauty}
                                      & HR@5    & 0.0413 & 0.0278 & 0.0500 & 0.0425 & 0.0560 & 0.0626 & 0.0608 & 0.0620 & 0.0601 & \underline{0.0627} & 0.0619 & 0.0623 & \bf{0.0684} &  9.16\% \\
                                      & HR@10   & 0.0646 & 0.0456 & 0.0709 & 0.0636 & 0.0788 & 0.0872 & 0.0857 & 0.0880 & 0.0849 & 0.0878 & \underline{0.0882} & 0.0879 & \bf{0.0948} &  7.48\% \\
                                      & HR@20   & 0.0959 & 0.0706 & 0.0966 & 0.0929 & 0.1090 & 0.1196 & 0.1180 & 0.1206 & 0.1167 & 0.1193 & \underline{0.1207} & 0.1204 & \bf{0.1311} &  8.62\% \\
                                      & NDCG@5  & 0.0267 & 0.0181 & 0.0350 & 0.0287 & 0.0392 & 0.0437 & 0.0435 & 0.0438 & 0.0426 & 0.0422 & 0.0438 & \underline{0.0441} & \bf{0.0488} & 10.57\% \\
                                      & NDCG@10 & 0.0342 & 0.0238 & 0.0418 & 0.0355 & 0.0465 & 0.0516 & 0.0514 & 0.0523 & 0.0504 & 0.0499 & 0.0519 & \underline{0.0524} & \bf{0.0573} &  9.39\% \\
                                      & NDCG@20 & 0.0421 & 0.0301 & 0.0483 & 0.0428 & 0.0541 & 0.0598 & 0.0595 & \underline{0.0605} & 0.0584 & 0.0600 & 0.0604 & \underline{0.0605} & \bf{0.0664} &  9.70\% \\
        \hline
        \multirow{6}{*}{Games}
                                      & HR@5    & 0.0541 & 0.0397 & 0.0659 & 0.0671 & 0.0796 & 0.0845 & 0.0846 & 0.0839 & 0.0857 & 0.0811 & 0.0863 & \underline{0.0870} & \bf{0.0934} &  7.35\% \\
                                      & HR@10   & 0.0853 & 0.0673 & 0.1019 & 0.1074 & 0.1218 & 0.1286 & 0.1263 & 0.1289 & 0.1277 & 0.1260 & 0.1310 & \underline{0.1323} & \bf{0.1423} &  7.55\% \\
                                      & HR@20   & 0.1291 & 0.1095 & 0.1485 & 0.1637 & 0.1753 & 0.1886 & 0.1827 & 0.1887 & 0.1848 & 0.1833 & 0.1879 & \underline{0.1894} & \bf{0.2073} &  9.43\% \\
                                      & NDCG@5  & 0.0345 & 0.0259 & 0.0437 & 0.0437 & 0.0527 & 0.0567 & 0.0569 & 0.0568 & 0.0575 & 0.0524 & 0.0575 & \underline{0.0579} & \bf{0.0629} &  8.62\% \\
                                      & NDCG@10 & 0.0445 & 0.0347 & 0.0553 & 0.0568 & 0.0662 & 0.0709 & 0.0703 & 0.0713 & 0.0710 & 0.0668 & 0.0718 & \underline{0.0725} & \bf{0.0786} &  8.48\% \\
                                      & NDCG@20 & 0.0555 & 0.0453 & 0.0671 & 0.0708 & 0.0798 & 0.0859 & 0.0845 & 0.0863 & 0.0854 & 0.0829 & 0.0861 & \underline{0.0868} & \bf{0.0950} &  9.43\% \\
        \hline
        \multirow{6}{*}{Sports}
                                      & HR@5    & 0.0224 & 0.0137 & 0.0278 & 0.0211 & 0.0317 & 0.0347 & 0.0332 & 0.0335 & 0.0348 & 0.0310 & 0.0349 & \underline{0.0360} & \bf{0.0379} &  5.32\% \\
                                      & HR@10   & 0.0352 & 0.0222 & 0.0395 & 0.0336 & 0.0462 & 0.0510 & 0.0470 & 0.0498 & 0.0511 & 0.0495 & 0.0516 & \underline{0.0527} & \bf{0.0550} &  4.42\% \\
                                      & HR@20   & 0.0551 & 0.0366 & 0.0560 & 0.0521 & 0.0666 & 0.0721 & 0.0665 & 0.0713 & 0.0723 & 0.0703 & 0.0727 & \underline{0.0736} & \bf{0.0796} &  8.12\% \\
                                      & NDCG@5  & 0.0144 & 0.0086 & 0.0197 & 0.0134 & 0.0217 & 0.0242 & 0.0239 & 0.0234 & 0.0245 & 0.0208 & 0.0247 & \underline{0.0252} & \bf{0.0263} &  4.28\% \\
                                      & NDCG@10 & 0.0185 & 0.0114 & 0.0236 & 0.0175 & 0.0263 & 0.0294 & 0.0283 & 0.0286 & 0.0298 & 0.0266 & 0.0298 & \underline{0.0306} & \bf{0.0318} &  4.07\% \\
                                      & NDCG@20 & 0.0235 & 0.0150 & 0.0277 & 0.0221 & 0.0315 & 0.0347 & 0.0333 & 0.0340 & 0.0351 & 0.0341 & 0.0352 & \underline{0.0359} & \bf{0.0380} &  5.88\% \\
        \hline
        \multirow{6}{*}{Yelp}
                                      & HR@5    & 0.0234 & 0.0243 & 0.0268 & 0.0319 & 0.0341 & 0.0374 & 0.0358 & \underline{0.0377} & 0.0371 & 0.0367 & 0.0364 & 0.0362 & \bf{0.0402} &  6.54\% \\
                                      & HR@10   & 0.0407 & 0.0385 & 0.0426 & 0.0527 & 0.0543 & \underline{0.0602} & 0.0582 & 0.0593 & 0.0596 & 0.0592 & 0.0587 & 0.0576 & \bf{0.0657} &  9.07\% \\
                                      & HR@20   & 0.0677 & 0.0598 & 0.0688 & 0.0865 & 0.0873 & 0.0975 & 0.0924 & 0.0949 & \underline{0.0984} & 0.0942 & 0.0938 & 0.0925 & \bf{0.1036} &  5.33\% \\
                                      & NDCG@5  & 0.0146 & 0.0152 & 0.0176 & 0.0208 & 0.0229 & \underline{0.0245} & 0.0235 & \underline{0.0245} & 0.0243 & 0.0240 & 0.0242 & 0.0243 & \bf{0.0268} &  9.21\% \\
                                      & NDCG@10 & 0.0202 & 0.0198 & 0.0227 & 0.0274 & 0.0293 & \underline{0.0318} & 0.0307 & 0.0314 & 0.0315 & 0.0311 & 0.0309 & 0.0312 & \bf{0.0350} & 10.01\% \\
                                      & NDCG@20 & 0.0269 & 0.0251 & 0.0293 & 0.0359 & 0.0375 & 0.0412 & 0.0392 & 0.0404 & \underline{0.0413} & 0.0400 & 0.0409 & 0.0400 & \bf{0.0445} &  7.69\% \\
        \hline
        \multirow{6}{*}{MovieLens 1M}
                                      & HR@5    & 0.1228 & 0.1061 & 0.1280 & 0.1416 & 0.1421 & 0.1581 & 0.1573 & 0.1599 & 0.1591 & 0.1465 & 0.1597 & \underline{0.1610} & \bf{0.1755} &  8.99\% \\
                                      & HR@10   & 0.1875 & 0.1539 & 0.1955 & 0.2197 & 0.2151 & 0.2346 & 0.2386 & 0.2346 & 0.2344 & 0.2314 & 0.2389 & \underline{0.2408} & \bf{0.2626} &  8.99\% \\
                                      & HR@20   & 0.2663 & 0.2142 & 0.2773 & 0.3171 & 0.3075 & 0.3307 & 0.3314 & 0.3291 & 0.3325 & 0.3261 & 0.3392 & \underline{0.3410} & \bf{0.3705} &  8.67\% \\
                                      & NDCG@5  & 0.0813 & 0.0704 & 0.0836 & 0.0914 & 0.0932 & 0.1031 & 0.1036 & 0.1053 & 0.1053 & 0.0947 & 0.1057 & \underline{0.1068} & \bf{0.1163} &  8.90\% \\
                                      & NDCG@10 & 0.0982 & 0.0857 & 0.1055 & 0.1164 & 0.1165 & 0.1277 & 0.1299 & 0.1295 & 0.1296 & 0.1220 & 0.1313 & \underline{0.1325} & \bf{0.1443} &  8.90\% \\
                                      & NDCG@20 & 0.1220 & 0.1009 & 0.1260 & 0.1409 & 0.1398 & 0.1521 & 0.1534 & 0.1533 & 0.1543 & 0.1497 & 0.1564 & \underline{0.1577} & \bf{0.1714} &  8.67\% \\
        \bottomrule
    \end{tabular}
\end{table*}

We conduct experiments on six publicly available datasets spanning diverse domains and scales, with detailed statistics summarized in \autoref{tab/dataset}.
(1) Following \cite{kang2018self, xie2022contrastive, du2023ensemble}, we use four Amazon \cite{mcauley2015image} product categories-\textit{Toys}, \textit{Beauty}, \textit{Games}, and \textit{Sports}.
(2) For Yelp \cite{asghar2016yelp}, we extract interactions collected after January 1st, 2020.
(3) From MovieLens \cite{harper2015movielens} collection, we adopt 1M version following \cite{sun2019bert4rec, xie2022contrastive}.
In line with prior works \cite{kang2018self, qiu2022contrastive}, all interactions are treated as implicit feedback, and users and items with fewer than five interactions are filtered out.

\subsubsection{Baseline}
We compare FLAME with three groups of methods:
\begin{itemize}[itemsep=1pt, left=0pt]
    \item \textbf{General Sequential Recommender}:
    Sequential recommendation models have been build on a wide range of architectures.
    GRU4Rec \cite{hidasi2015session} employs RNNs to model user sequences.
    Caser \cite{tang2018personalized} applies CNNs to capture local patterns.
    SASRec \cite{kang2018self} leverages self-attention to dynamically identify item relevance.
    BERT4Rec \cite{sun2019bert4rec} adopts bidirectional Transformer with masked item prediction.
    FMLPRec \cite{zhou2022filter} replaces attention block of SASRec with feedforward MLP for frequency-based modeling.
    \item \textbf{Contrastive Learning Sequential Recommender}:
    These methods create multiple sequence representations from a single network by employing various view-generation techniques for contrastive optimization.
    CL4SRec \cite{xie2022contrastive} applies data-level augmentations.
    DuoRec \cite{qiu2022contrastive} treats same target sequences as positive pairs.
    MCLRec \cite{qin2023meta} combines both data-level and model-level augmentations.
    CT4Rec \cite{chong2023ct4rec} introduces dual contrastive objectives.
    \item \textbf{Ensemble-based Sequential Recommender}:
    EMKD \cite{du2023ensemble} ensembles multiple BERT4Rec networks with contrastive distillation.
    DHEN \cite{zhang2022dhen} hierarchically models diverse interaction patterns with heterogeneous architectures.
    FamouSRec \cite{zhang2025frequency} integrates heterogeneous architectures with frequency-aware contrastive learning.
\end{itemize}

\subsubsection{Evaluation}
We adopt the standard leave-one-out protocol \cite{kang2018self}, where for each user sequence, the last item is used for testing and the penultimate item for validation.
To ensure fairness, we perform a full-ranking evaluation without negative sampling \cite{qiu2022contrastive} and report performance using HR@$K$ and NDCG@$K$ for $K \in \{ 5, 10, 20 \}$.

\subsubsection{Implementation}
FLAME and all baselines are implemented in PyTorch.
For fair comparison, we thoroughly tune each baseline according to its original paper.
For EMKD, we ensemble four BERT4Recs.
For DHEN, we empicially find that self-attention and MLP pairing is the most effective combination.
For FamouSRec, we ensembles four heterogeneous architectures: self-attention, GRU, MLP, and Mamba.
In FLAME, the Transformer encoder consists of $L = 2$ layers and 2 attention heads, with a dropout rate of 0.5.
The embedding dimension $d$ is 64, and the maximum sequence length $T$ is 50.
As described in Section~\ref{sec/ensemble_construction}, we decompose both the \textit{Frozen} and \textit{Learnable} networks into $M = 2$ sub-modules by default.
For the contrastive alignment loss, the temperature $\tau$ is selected from \{0.1, 0.5, 1, 5, 10\}, and the initial loss coefficient $\lambda_0$ from \{1e-4, 1e-3, 1e-2, 1e-1, 1e-0\}, with a fixed terminal loss coefficient $\lambda_R = 1e-5$.
All models are trained for $R = 200$ epochs using Adam \cite{kingma2014adam} optimizer with a learning rate of 0.001 and batch size of 256.
We apply early stopping based on validation NDCG@20 with a patience of 30 epochs.
All experiments are conducted on a single NVIDIA GeForce RTX 3090 GPU.

\subsection{Performance Comparison (RQ1)} \label{sec/performance_comparison}

To assess the effectiveness of FLAME in sequential recommendation, we conduct extensive experiments on six datasets, whose results are summarized in \autoref{tab/performance}.
Across all datasets and evaluation metrics, FLAME consistently outperforms all baselines, achieving relative improvements ranging from 4.07\% to 10.57\%.
Beyond this strong performance, we highlight three key observations that provide deeper insight into the comparative performance:
\begin{itemize}[itemsep=1pt, left=0pt]
    \item General sequential recommenders exhibit overall inferior performance, suggesting that a single network struggles to capture the diverse user behavior patterns and item characteristics in sequential recommendation data.
    \item Contrastive learning models outperform general baselines, indicating the effectiveness of contrastive objectives and supporting the validity of the representation alignment strategy of FLAME.
    Despite their improvements, however, these methods still operate within a single-network paradigm, limiting their ability to model representation diversity.
    \item Ensembles like FamouSRec use multiple networks to achieve strong performance through diversity, but often suffer from training instability caused by noisy mutual supervision (see Section~\ref{sec/efficiency_and_stability_study}).
    In contrast, FLAME addresses this issue by incorporating a pretrained and frozen network into modular ensemble design, stabilizing training while preserving diversity.
\end{itemize}

\subsection{Ablation Study (RQ2)} \label{sec/ablation_study}

\begin{table}[t]
    \footnotesize
    \setlength{\tabcolsep}{3pt}
    \caption{Ablation study with crucial components.}
    \label{tab/ablation_study}
    \begin{tabular}{l|cc|cc}
        \toprule
        \multirow{2}{*}{\textbf{Model}} & \multicolumn{2}{c|}{\textbf{Beauty}} & \multicolumn{2}{c}{\textbf{MovieLens 1M}} \\
        & HR@20 & NDCG@20 & HR@20 & NDCG@20 \\
        \midrule
        (1) FLAME              & \bf{0.1311} & \bf{0.0664} & \bf{0.3705} & \bf{0.1714} \\
        \hline
        (2) w/o modular        &     0.1265  &     0.0618  &     0.3549  &     0.1603  \\
        (3) w/o guide          &     0.1254  &     0.0618  &     0.3446  &     0.1576  \\
        \hline
        (4) w/o weighting      &     0.1275  &     0.0645  &     0.3586  &     0.1650  \\
        (5) w/o annealing      &     0.1294  &     0.0639  &     0.3575  &     0.1653  \\
        \bottomrule
    \end{tabular}
\end{table}

To evaluate the effectiveness of individual components in the proposed FLAME framework, we conduct an ablation study by comparing the default implementation against several controlled variants.
Specifically, we define the following five settings:
\begin{itemize}[itemsep=1pt, left=0pt]
    \item \textbf{(1) FLAME} serves as the full model.
    \item \textbf{(2) w/o modular} adopts conventional ensemble approach.
    \item \textbf{(3) w/o guide} trains both networks from scratch.
    \item \textbf{(4) w/o weighting} replaces $\mathcal{L}_{mkt}$ with $\mathcal{L}_{cl}$.
    \item \textbf{(5) w/o annealing} employs a constant loss coefficient $\lambda = \lambda_0$.
\end{itemize}

Presenting the performance of the aforementioned FLAME variants in \autoref{tab/ablation_study}, we draw two key findings.
First, the substantial performance drop (up to 8.76\%) without modular ensemble (2) or guided mutual learning (3) confirms their critical role in building diverse and reliable representations.
Second, the degraded performance without similarity-based weighting (4) or coefficient annealing (5) suggests their meaningful contribution to representation alignment and training stability.

\subsection{Scalability Analysis (RQ3)} \label{sec/scalability_analysis}

\begin{figure}[t]
  \centering
  \begin{subfigure}[t]{0.23\textwidth}
    \centering
    \includegraphics[width=\textwidth]{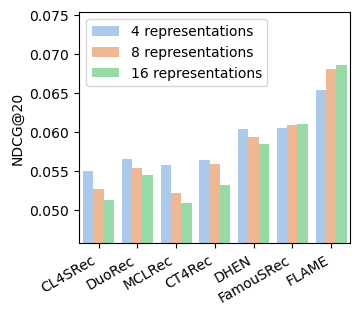}
    \caption{Beauty}
  \end{subfigure}
  \hfill
  \begin{subfigure}[t]{0.23\textwidth}
    \centering
    \includegraphics[width=\textwidth]{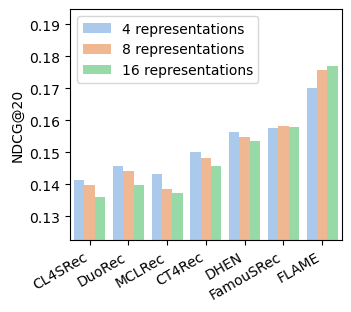}
    \caption{MovieLens 1M}
  \end{subfigure}
  \caption{Model performance of FLAME and baselines with varying number of representations.}
  \label{fig/flame_granularity}
  \Description{}
\end{figure}

In this section, we investigate how FLAME behaves under a finer-grained decomposition scheme.
Beyond the default two sub-modules setup, we further perform a layer-wise decomposition of the sequence encoder within each network, diving a network with an $L$-layer encoder into $M = L + 1$ sub-modules.

To determine if the performance gain of FLAME stems from genuine ensemble diversity rather than merely an increased number of vies, we conduct a controlled experiment.
We evaluate FLAME with $L = 1, 2, 3$ (yielding 4, 8, and 16 representations, respectively) against baselines that are adjusted to produce the same number of views\footnote{Contrastive learning baselines repeat their augmentation strategies to match the number of views. DHEN and FamouSRec multiplies the number of layers and networks, respectively. EMKD is excluded since it already employs more than $2^4 = 16$ representations.}, with the comparison shown in \autoref{fig/flame_granularity}.
Through the comparison, we make the following two observations:
\begin{itemize}[itemsep=1pt, left=0pt]
    \item Contrastive learning baselines degrade even with more representations, likely due to limited diversity from a single network.
    In contrast, FLAME improves with finer-grained sub-module decomposition, as modular ensemble enables structural diversity, thereby enhancing recommendation performance.
    \item Heterogeneous ensembles like DHEN and FamouSRec exhibit poor scalability, as their reliance on architectural differences for diversity imposes structural constraints that hinder effective knowledge sharing, leading to marginal or even negative gains despite linearly increasing costs.
    In contrast, modular ensemble in FLAME achieves high diversity within homogeneous networks without these limitations, efficiently boosting performance by up to 4.74\% at no additional model capacity and striking a superior performance-efficiency balance.
\end{itemize}

\begin{figure}[t]
  \centering
  \begin{subfigure}[t]{0.23\textwidth}
    \centering
    \includegraphics[width=\textwidth]{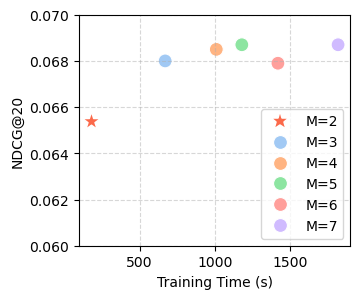}
    \caption{Beauty}
  \end{subfigure}
  \hfill
  \begin{subfigure}[t]{0.23\textwidth}
    \centering
    \includegraphics[width=\textwidth]{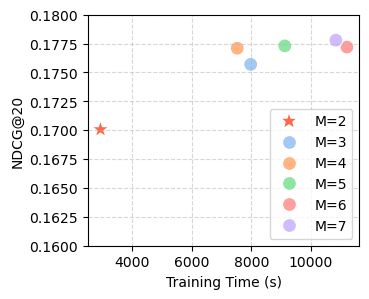}
    \caption{MovieLens 1M}
  \end{subfigure}
  \caption{Comparison between training wall clock time and model performance of FLAME with varying number of decomposed sub-modules $M$.}
  \label{fig/flame_granularity2}
  \Description{}
\end{figure}

We further analyze the trade-off between performance and efficiency in FLAME by expanding the number of sub-modules $M$ from 2 to 7.
As illustrated in \autoref{fig/flame_granularity2}, increasing the number of sub-modules $M$ leads to a substantial increase in training time, while performance gains become marginal beyond $M = 4$.
This diminishing return is attributed to the escalating difficulty of aligning a vast number of modular representations into a unified semantic space.
Therefore, although FLAME can achieve slightly better performance with larger $M$ values, we conclude that our default setting of $M = 2$ offers the most reasonable and practical trade-off between training efficiency and model performance.

\subsection{Frozen Network Robustness (RQ4)} \label{sec/frozen_network_robustness}

\begin{figure}[t]
  \centering
  \begin{subfigure}[t]{0.23\textwidth}
    \centering
    \includegraphics[width=\textwidth]{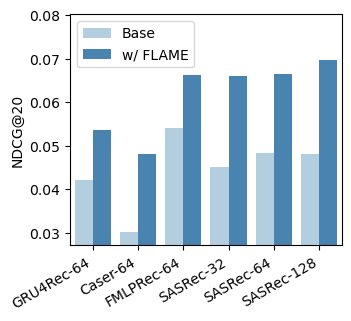}
    \caption{Beauty}
  \end{subfigure}
  \hfill
  \begin{subfigure}[t]{0.23\textwidth}
    \centering
    \includegraphics[width=\textwidth]{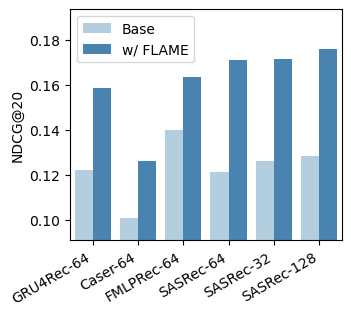}
    \caption{MovieLens 1M}
  \end{subfigure}
  \caption{Model performance of applying FLAME across different {\it Frozen} network with varying architectures and dimensionalities $d$.}
  \label{fig/flame_robustness}
  \Description{}
\end{figure}

In this section, we investigate the robustness of FLAME regarding the selection of the \textit{Frozen} network, presenting the performance variations of FLAME in \autoref{fig/flame_robustness}.

\textbf{Architecture-agnostic Applicability.}
In the default setup ($M = 2$), we decompose both the \textit{Frozen} and \textit{Learnable} networks into an embedding module and a sequence encoding module.
This decomposition strategy is prevalent throughout sequential recommendation literature, rendering FLAME technically adaptable to diverse architectures.
To evaluate this applicability, We apply FLAME to four representative models: GRU4Rec (RNN), Caser (CNN), FMLPRec (MLP), and SASRec (Transformer).
FLAME consistently improves performance across all backbones, demonstrating that the framework effectively condenses ensemble diversity regardless of the underlying sequence modeling mechanism.

\textbf{Capacity Sensitivity.}
We further analyze the impact of model capacity by varying the dimensionality $d$.
While SASRec-32 performs poorly compared to SASRec-64, FLAME-32 matches the performance of FLAME-64, confirming robustness against weaker teachers.
Conversely, while SASRec-128 saturates showing negligible gains, FLAME-128 achieves clear improvement over FLAME-64.
This indicates that FLAME effectively capitalizes on increased capacity, leveraging the expanded representation space through modular ensemble and guided mutual learning.

\subsection{Efficiency and Stability Study (RQ5)} \label{sec/efficiency_and_stability_study}

In this section, we rigorously examine the ability of FLAME to overcome two major limitations of conventional ensemble methods—\textit{computational inefficiency} and \textit{training instability}.

\textbf{Computational Efficiency.}
\begin{table}[t]
    \footnotesize
    \setlength{\tabcolsep}{2pt}
    \caption{Model performance and inference latency with different number of parameters. SASRec-Large doubles the embedding size $d$ of SASRec. SASRec-Ensemble corrsponds to Ensemble-Scratch in \autoref{fig/ensemble_instability}.}
    \label{tab/flame_efficiency}
    \begin{tabular}{l|ccc|ccc}
        \toprule
        \multirow{2}{*}{\textbf{Model}} & \multicolumn{3}{c|}{\textbf{Beauty}} & \multicolumn{3}{c}{\textbf{MovieLens 1M}} \\
        & \# Params. & Infer. & NDCG@20 & \# Params. & Infer. & NDCG@20 \\
        \midrule
        SASRec          & 0.84 M & \bf{1.30 s} &     0.0483  & 0.31 M & \bf{0.16 s} &     0.1260  \\
        SASRec-Large    & 1.86 M &     1.35 s  &     0.0485  & 0.80 M &     0.18 s  &     0.1227  \\
        SASRec-Ensemble & 1.67 M &     2.58 s  &     0.0513  & 0.61 M &     0.29 s  &     0.1352  \\
        \hline
        EMKD            & 4.76 M &     5.52 s  &     0.0600  & 1.56 M &     1.08 s  &     0.1497  \\
        DHEN            & 3.12 M &     2.97 s  &     0.0604  & 1.49 M &     0.44 s  &     0.1564  \\
        FamouSRec       & 2.30 M &     3.29 s  &     0.0605  & 1.77 M &     0.49 s  &     0.1577  \\
        \hline
        FLAME           & 1.67 M & \bf{1.30 s} & \bf{0.0660} & 0.61 M & \bf{0.16 s} & \bf{0.1714} \\
        \bottomrule
    \end{tabular}
\end{table}

We evaluate the efficiency of FLAME by comparing the performance and inference latency of various baselines relative to their parameter count, as shown in \autoref{tab/flame_efficiency}.
Our results reveal that simply increasing model capacity is ineffective and can even harm performances (SASRec-Large), whereas ensembling is a more parameter-efficient strategy for capturing diversity (SASRec-Ensemble).
While advanced ensembles like EMKD, DHEN, and FamouSRec further boost performance, they do so at the cost of significant parameter and latency.
In sharp contrast, FLAME uses modular ensemble to attain state-of-the-art performance with significantly fewer parameters.
For inference, it uses only the \textit{Learnable} network, matching the deployment size of single SASRec and achieving up to 6.75 times lower latency than other conventional ensembles.

\textbf{Training Stability.}
\begin{figure}[t]
  \centering
  \begin{subfigure}[t]{0.23\textwidth}
    \centering
    \includegraphics[width=\textwidth]{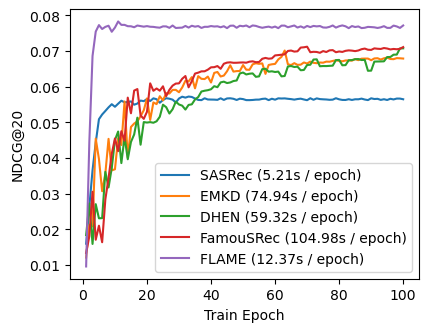}
    \caption{Beauty}
  \end{subfigure}
  \hfill
  \begin{subfigure}[t]{0.23\textwidth}
    \centering
    \includegraphics[width=\textwidth]{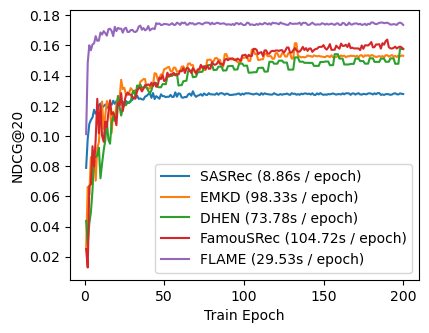}
    \caption{MovieLens 1M}
  \end{subfigure}
  \caption{Training curves with per-epoch latency for SASRec, EMKD, DHEN, FamouSRec and FLAME.}
  \label{fig/flame_stability}
  \Description{}
\end{figure}

As the learning curves in \autoref{fig/flame_stability} show, conventional ensembles like EMKD, DHEN and FamouSRec exhibit unstable convergence, especially during early training phases, and their learning progress is even slower than that of SASRec.
This highlights the adverse effect of noisy mutual supervision, which is further exacerbated in sparse datasets such as Beauty.
In contrast, FLAME not only converges more stably but also faster than SASRec, indicating that guided mutual learning via the \textit{Frozen} network effectively mitigates the instability inherent in conventional ensemble training.
Concretely, FLAME achieves 4.55 to 7.69 times faster convergence than the other ensemble-based baselines.

\subsection{Parameter Sensitivity (RQ6)} \label{sec/parameter_sensitivity}
\begin{figure}[t]
  \centering
  \includegraphics[width=0.23\textwidth]{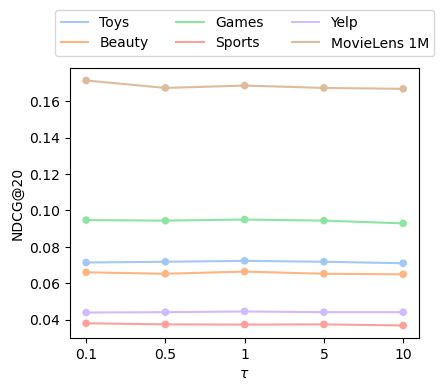}
  \includegraphics[width=0.23\textwidth]{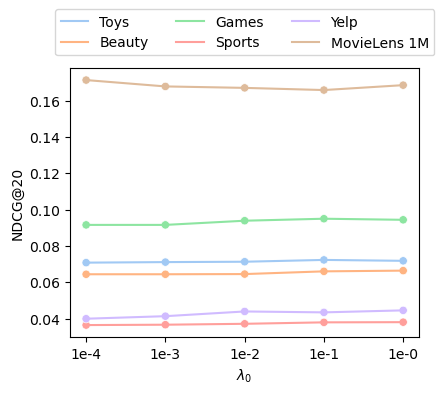}
  \caption{Parameter sensitivity of $\tau$ and $\lambda_0$.}
  \label{fig/flame_parameter}
  \Description{}
\end{figure}

Another strength of FLAME lies in its low sensitivity to hyperparameters.
Without modifying the backbone architecture, the fusion of modular ensemble and guided mutual learning in FLAME effectively stabilizes training while achieving ensemble-level diversity.
To further assess this robustness, we conduct sensitivity experiments on two hyperparameters in contrastive alignment loss: temperature $\tau$ and initial loss coefficient $\lambda_0$.
Although these are not core design parameters of FLAME, they appear in the contrastive objective and can influence the training process.
The results in \autoref{fig/flame_parameter} show that FLAME consistently maintains strong performance across a wide range of settings, suggesting that the framework is robust even under varying contrastive configurations.

\section{Conclusion}
We present FLAME, a novel framework achieving the benefits of ensemble learning--diversity and reliability--without the associated high costs.
FLAME utilizes just two networks, one frozen and one learnable, to efficiently generate diverse representations via sub-module combinations, while the frozen network provides stable guidance.
Crucially, this design condenses ensemble power into a single learnable network, ensuring zero-overhead inference suitable for real-time applications.
Consequently, FLAME trains faster and more stably while achieving stronger performance, suggesting that scalable and effective ensemble learning is attainable without typical trade-offs.
Future work could investigate elastic inference mechanisms that dynamically activate ensemble paths for high-uncertainty users to further optimize the accuracy-efficiency trade-off.

\begin{acks}
    This work was supported by the NRF grant funded by the MSIT (No. RS-2024-00335873), the IITP grant funded by the MSIT (No.RS-2019-II191906, Artificial Intelligence Graduate School Program(POSTECH)), and the INNOPOLIS grant funded by the MSIT (No. RS-2025-25449754).
    This work was also supported by ICT Creative Consilience Program through the IITP grant funded by the MSIT (IITP-2026-RS-2020-II201819) and Basic Science Research Program through the NRF funded by the Ministry of Education (NRF-2021R1A6A1A03045425).
\end{acks}

\bibliographystyle{ACM-Reference-Format}
\balance
\bibliography{references}

\end{document}